\documentclass[10pt,fleqn]{article}
\usepackage{amssymb}

\usepackage{amsmath}

\usepackage{amsmath,amssymb,amsthm}

\begin{document}
{\Large
\begin{center}
{\Large \textbf{Differential Invariants and Hidden
Symmetry}}

\vskip 20pt {\large \textbf{Irina YEHORCHENKO}}

\vskip 20pt {Institute of Mathematics of NAS Ukraine, 3
Tereshchenkivs'ka Str., 01601 Kyiv-4, Ukraine} \\
E-mail: iyegorch@imath.kiev.ua
\end{center}

\vskip 50pt
\begin{abstract}
We show an algorithm for description of classes of equations having
specific conditional or hidden symmetry, and/or reducible with a
specific ansatz. We consider reductions that are due to Lie, conditional
and Type II hidden symmetry. We also discuss relations
between the concepts of hidden and conditional symmetry.
As examples, we describe general classes equations having hidden and conditional symmetry
under rotations and boosts in the Lorentz and Euclid groups.
\end{abstract}

\section{Background Concepts}

One of the key problems within the field of symmetry analysis of
differential equations is description
of equations with particular pre-determined symmetry properties.
Choosing equations with particular symmetry properties may mean that we
have equations with solutions of some particular structure.

It was proved in \cite{zhdanov&tsyfra&popovych99} that reduction of PDE
(direct ansatz approach) is equivalent to the non-classical (conditional) symmetry.
Thus, the conditional
invariance of a differential equation under an involutive family
of first-order differential operators $Q_a$ is equivalent to
possibility of reduction of this equation by means of the ansatz
corresponding to this family of operators.

So, description of all equations from some class having a particular conditional symmetry
(we will deal, speaking more rigourously,
with $Q$-conditional symmetry, as defined in \cite{FSS}) will give all equations from this
class that can be reduced by means of the specific ansatz corresponding to conditional symmetry operators
being considered.

\noindent {\bf Definition 1.} {\it The equation
$F(x,u,\underset{1}{u},\ldots , \underset{l}{u})=0$, where
$\underset{k}{u}$ is the set of all $k$th-order partial
derivatives of the function $u=(u^1,u^2,\ldots ,u^m)$, is called
$Q$-conditionally invariant \cite{FSS} under the operator
\[
Q=\xi ^i(x,u)\partial_{x_i}+\eta ^r(x,u)\partial_{u^r}\nonumber
\]
if there is an additional condition
\begin{equation}
Qu=0, \label{G=0}
\end{equation}
such that the system of two equations $F=0$, $Qu=O$ is
invariant under the operator $Q$}. All differential consequences
of the condition $Qu=0$ shall be taken into account up to the
order $l-1$.

Note that we will need such definition of conditional symmetry
if we specifically want to describe equations reducible with some
form of ansatz; however, we can use a more general definition of
conditional symmetry and describe classes of equations having such
symmetry.

\noindent {\bf Definition 2.} {\it The equation
$F(x,u,\underset{1}{u},\ldots , \underset{l}{u})=0$, where
$\underset{k}{u}$ is the set of all $k$th-order partial
derivatives of the function $u=(u^1,u^2,\ldots ,u^m)$, is called
conditionally invariant \cite{FSS},
if there is an additional condition
$G(x,u,\underset{1}{u},\ldots , \underset{l_1}{u})=0$
such that the system of two equations $F=0$, $G=0$ is
invariant under some operator
\[
Q=\xi ^i(x,u)\partial_{x_i}+\eta ^r(x,u)\partial_{u^r}\nonumber
\]
\noindent
that is not a Lie invariance operator of the equation $F=0$.
All differential consequences
of the condition $G=0$ shall be taken into account up to the
order $l-l_1$.}

In this paper we give an outlook of description of partial
differential equations possessing certain conditional and
hidden symmetries.

Group classification for classes of differential equations is
aimed at identification of equations having wider symmetries than
the equations of the class in general. For an overview of the
group classification problems and the extensive list of related
references see \cite{PopovychIvanova}. Usually two
types of such problems are considered -- finding the equations
within a general class that are invariant under specific symmetry
group, and description of all symmetries (up to appropriately
chosen equivalence) of equations that belong to the specific
class. On the basis of the known algorithms for group
classification of differential equations in the Lie's sense we
develop approaches for a~systematic description of classes of
nonlinear PDEs that display conditional and hidden symmetry.

The definition of conditional differential invariants presented
below refers to the Definition of the conditional symmetry.

\noindent 
{\bf Definition 3.} {\it A class of equations can be regarded as
general if any local transformations of dependent and
independent variables transform the equations into an equation
within the same class, e.g.\ the class of all $k$th-order PDEs
$F=F(x,u,\underset{1}{u},\ldots ,\underset{k}{u})=0$ with $x$, $u$
being respectively $n$- and $m$-dimensional independent and
dependent variables, $\underset{r}{u}$ being the set of all
$r$-th-order partial derivatives of the function
$u=\big(u^1,u^2,\ldots ,u^m\big)$.}

Group classification even with respect to the Lie symmetry for the
general classes is usually an overwhelming task, and, to our
knowledge, such problem was completely solved  only for single
ordinary differential equations by S.~Lie
\cite{Lie1881withTrans}. A restricted, but practically
important problem for the general classes of equations would be
description of all equations within the class invariant under some
specified symmetry group that can be done by describing all
differential invariants for such group. Similarly, description of
equations having specified Lie symmetry and specified conditional
symmetry may be done by means of conditional differential
invariants, as shown in \cite{cond-diff-inv}.

For a more specific class, it may be possible to perform a full group
classification of the system that consists of the original
equation together with the reduction conditions of the type $Q_a
[u]=0$ (with appropriate prolongations of the conditions).

\noindent {\bf Definition 4.} \cite{cond-diff-inv} {\it A function
$F(x,u,{\mathop {u}\limits_1}, \ldots, {\mathop {u}\limits_k})$ is
a conditional differential invariant of the operator $Q$, if
under the conditions $G(x,u,\underset{1}{u},\ldots,
\underset{r}{u})=0$ the relations $Q[F]=0$, $Q[G]=0$ are
satisfied. We take prolongations of the operators of the order
$\max(k,r)$.}

A set of invariants of the order $r \le k$ of the operator $Q$
with the conditions $G=0$  is called a {\it generating set} of the
$k$th-order conditional differential invariants of the algebra $Q$
if all other invariants can be represented as functions of
invariants in this set.

Invariants in such generating set may be both absolute invariants
of $Q$ and $G$-conditional of the form $G^{(l)}\times R_{(l)}$,
where $G^{(l)}$ are derivatives of $G$ of the order $l \le k-r$
and $R_{(l)}$ are arbitrary functions determined on the manifold
$G^{(k)}=0$ for all values of $k$.

The number of functionally independent $Q$-absolute invariants in the
generating set of conditional differential invariants can be
calculated similarly to the number of invariants in a functional
basis of absolute differential invariants, as $s-1$, where $s$ is
the number of variables in  the set ${x,u,{\mathop {u}\limits_1},
\ldots, {\mathop {u}\limits_k}}$. Number of independent
purely $G$-conditional invariants is equal to the number of
independent conditions of the type $G^{(l)}=0$ and their differential
consequences.

In some cases we would be able to construct a functional basis of
conditional invariants, i.e. the maximal set of functionally
independent conditional invariants. That is possible e.g. in the case
when we put a requirement that our conditional invariants should
be also absolute invariants of some Lie algebra $L$, and
additional conditions $G=0$ in the Definition 3 and their relevant
differential consequences are not invariant under $L$.

\section{Relation of Hidden Symmetry and Conditional Symmetry}
Further we will treat Type II hidden symmetry as a partial case of conditional
symmetry, and discuss a systematic approach to description of equations with some
specific hidden symmetry, or equations that may have such symmetries, within the lines
of such conditional symmetry approach.

A very close relation of the hidden symmetry (for some
initial papers on the subject see \cite{Abraham}) to
conditional symmetry (for definitions and examples see e.g. \cite{cond-refs}) is
well-known. The concept of ``hidden symmetry'' has quite a few different
meanings in various contexts, and it is usually a symmetry not
obtainable by some standard and straightforward procedure
applicable to the models in this context. This term shares the
usage of other related terms like ``conditional symmetry'',
``approximate symmetry'', and ``symmetry'' or ``invariance'' when
the same words may denote rather different concepts.

Here we will consider hidden symmetry of partial differential equations
similarly to Type II hidden symmetry of ordinary differential
equations generally within the context of \cite{Abraham}. With respect to
ODE such symmetry arises as symmetry of equations with reduced
order that is not a symmetry of the original equations. In the
same way, for a PDE it is symmetry of the reduced equation (with
reduced number of independent variables) not present in the
original equation. However, we would like to point out that we
will consider all possible reductions to find hidden symmetries, not
only symmetry reductions.

\noindent {\bf Definition 5.} {\it A differential equation is said to have
hidden symmetry under an operator $X$ if after the process
of reduction of the number of independent variables the resulting
reduced equation is invariant under the operator $X_1$ (being the
projection of the operator $X$ in new variables) while the
original equation is not invariant under the operator $X$.}

Such symmetry may be "classical" in the sense that full hidden
symmetry of either ODE or PDE may be found by consecutive
symmetry reduction of the original equation and investigation of Lie
symmetries of the reduced equations, as provided by
L.~Ovsyannikov's {\it Submodels} programme
\cite{OvsyannikovSubmodels}. ``Symmetry'' or ``Lie
symmetry'' is determined in accordance to the procedures that may
be found e.g.\ in
\cite{{Ovs-eng},{books}}.

An interesting discussion of the origin and nature of the Type II hidden
symmetry was presented in \cite{AbrG08}. To see clearly the origin of hidden symmetries,
it is necessary always to keep in mind the additional conditions (representing Lie
or non-classical symmetries) that resulted in reduction to the new equations
having new symmetries that are hidden symmetries for an original equation.

The additional
symmetry under the operator
 $X_1$ of the reduced equation (hidden symmetry under the operator
 $X$ for the original equation)
 turns out to be a conditional symmetry of the original equation under
 conditions $Q_a [u]=0$ ($Q_a [u]$ designate characteristics of the
 vector field $Q_a$) with all appropriate differential consequences.
 Note that $X_1$ is a Lie symmetry of the
 reduced equation, and we do not add the condition $X[u]=0$ to the set of conditions,
 so such operator will not present a proper $Q$-conditional symmetry in the sense
 of~\cite{FSS}.

$Q$-conditional symmetry can also be hidden -- that is being a new
$Q$-conditional symmetry of the reduced equation. For examples of
such symmetries see e.g. \cite{YeVconf2003}.

When we look for equations with fully defined hidden symmetries, we can
describe such equations by means of conditional differential invariants -
both reduction conditions and hidden symmetry operators should be used as
conditions for such invariants.

An algorithm for group classification with respect to
hidden symmetry in the situation when hidden symmetries are not known
from the start:

Step 1. Obtain reduced equations for the class of initial PDE, using possible Lie
and conditional symmetries - this step requires a standard group classification
with respect to Lie and conditional symmetries.

Step 2. Perform group classification of the reduced equations. This
is the step when new symmetries of the reduced equations may be found
(and may be not - in this case the class of equations will
have no hidden symmetries). This step involves finding of equivalence
group of the class and of subclasses.

Step 3. Multiply inequivalent invariant reduced equations by means
of transformations from the equivalence group of this class (or subclasses).

Step 4. Go back to the original class of equations: find equations from the initial
class of PDE corresponding to the multiplied reduced equations.

Step 5. Find all inequivalent equations with respect to
transformations from the equivalence group of the initial class of
PDE.

A nontrivial hidden symmetry for partial differential
equations stems from the reduced equations having wider
equivalence group than the original equations (see \cite{Lisle-thesis}). So group classification
of the classes of equations with respect to hidden symmetry
involves study of equivalence groups of such classes,
in the similar way as it is done for classification with respect to the Lie symmetry.

``Simple'' hidden symmetries (Lie symmetries of reduced equations)
for a particular class of equations can be found by means of
consecutive Lie reductions and consecutive finding Lie symmetries
of the reduced equations. Group classification of such class with
respect to hidden symmetries of reduced equations will involve
description of all possible reductions and group classification of
the respective classes of reduced equations.

In \cite{HS2003} we considered a rather simple example of the nonlinear wave equation in two
spatial dimensions
\begin{equation}  \label{nl-wave1}
\Box u = f(t,x,y,u)
\end{equation}

Here we use the usual notations for partial derivatives and the d'Alembert operator.

Group classification of equation (\ref{nl-wave1}) with respect to
hidden symmetries for reduction by means of the
operator~$\partial_{y}$ was presented. Such reduction leads to the two-dimensional wave equation
$
u_{tt}-u_{xx} = f(t,x,u).
$
The next step is the usual group classification of the reduced
equation. It was given in \cite{Ovs-eng} ($f_{uu}=0$) and in
\cite{LM-nlwave} ($f_{uu}\ne 0$). Such group
classification was performed up to transformations from the
equivalence group of the equation.

Let us note that conditional symmetry of equation (\ref{nl-wave1}) was studied
 in \cite{CSPreprint2010}. "Extension" of dimensions of the found equations
with nontrivial conditional symmetries in this class will produce new multidimensional
equations with hidden symmetries.

\section{Example of a General Class: Hidden Symmetry with Respect to Translations}
For simplicity we use an example of a general class of all second-order
PDE for a scalar function $u$ and three independent variables $t,x,y$.
This class includes many physically interesting evolution and wave
equations. The ideas presented can be easily extended to equations
with larger number of dimensions.

Such class includes all equations of the form
\begin{equation}  \label{eq1}
F=F(t,x,y,u,\underset{1}{u},\underset{2}{u})=0.
\end{equation}

We will start with a straightforward example - description
of all such equations having Lie symmetry with respect to
the operator $\partial_x$ and hidden symmetry with respect to the
operator $\partial_y$ after reduction by means of the operator
$\partial_x$. The condition of such Lie and hidden symmetry,
according to Definition 1, is invariance of the
equation~(\ref{eq1}) under the operator $\partial_y$
on condition that $u_x=0$:
\begin{equation}  \label{eq1-invcond}
\partial_x F \big |_{F=0}=0, \qquad
\partial_y F \big |_{F=0, \ u_x=0}=0.
\end{equation}

The general solution of the
condition~(\ref{eq1-invcond}) will be a function of
all invariants of the operators $\partial_x$ and $\partial_y$,
that is $t$, $u$, $u_t$, $u_x$, $u_y$(being an absolute invariant of
$\partial_x$), and of the conditional invariants
\begin{gather}
q^1=u_xR^1(t,y,u,\underset{1}{u},\underset{2}{u}), \\
q^2=u_{xt}R^2(t,y,u,\underset{1}{u},\underset{2}{u}), \nonumber \\
q^3=u_{xx}R^3(t,y,u,\underset{1}{u},\underset{2}{u}), \nonumber \\
q^4=u_{xy}R^4(t,y,u,\underset{1}{u},\underset{2}{u}),
\end{gather}
\noindent 
where $R^k$ are arbitrary functions that is
reasonably determined on the relevant manifolds $u_x=0$, $u_{xt}=0$,
$u_{xx}=0$, $u_{xy}=0$:
\begin{equation}  \label{inv1}
F(q^k,t,u,\underset{1}{u},\underset{2}{u})=0,
\end{equation}

$F$ has to be a function of the invariants of the hidden symmetry
operator on the manifold determined by the reduction condition,
and have arbitrary form elsewhere. Note that the functions~$q^k$
in~(\ref{inv1}) are not entirely arbitrary: we cannot take e.g.\
\[
q^1=R^1(y,t,u,\underset{1}{u},\underset{2}{u})=u_x\frac{\widetilde{R}^1}{u_x}
\]
\noindent 
as such $R^1=\frac{\widetilde{R}^1}{u_x}$ will be in the general case
undetermined on the manifold $u_x=0$.

Equation~(\ref{inv1}) is reduced by means of the
operator $\partial_x$ to the equation
$$F_1(t,u,u_t,u_y,u_{tt},u_{ty},u_{yy})=0$$
that is invariant with respect to $\partial_y$. If $R^k_y \ne 0$ in
at least one expression for $q^k$ in (\ref{inv1}), then this
equation is not invariant with respect to the operator
$\partial_y$, and the relevant hidden symmetry is a proper hidden symmetry.

This example is easily generalised for larger order or equations
or number of reduction ope\-ra\-tors.

More specific examples of equations with hidden translational symmetry
are listed below:
\begin{gather}
u_t + u_xK_1(t,y,u)+ u_yK_2(t,u)+ u_{xx}+u_{yy}=0;\nonumber \\
u_{tt} - (K_1(t,y,u)u_x)_x - (K_2(t,u)u_y)_y=0.\nonumber
\end{gather}

These equations can be reduced by means of the operator $\partial_x$
to new equations invariant under $\partial_y$ - so they have hidden symmetry
with respect to the operator $\partial_y$.

\section{Equations Reducible using Radial Variables}

As it was shown e.g. in \cite{YeVconf2003} and then in
\cite{Abraham-wave}, reduction using radial variables often
results in the reduced equation with new symmetries as compared to the
initial equation, thus contributing to description of its hidden symmetries.

In this section we will describe all equations of the type (\ref{eq1})
that can be reduced by means of radial variables
\begin{gather}
r=x^2+y^2 \label{r}\\
\rho=t^2-x^2-y^2. \label{rho}
\end{gather}

Reduction of equation (\ref{eq1}) by means of the new variable (\ref{r})
is equivalent to its conditional invariance under the rotation operator
\begin{equation}
J=x\partial_y-y\partial_x. \label{rotation}
\end{equation}

Conditional differential invariants with the condition
\begin{equation}
xu_y-yu_x=0. \label{cond-rotation}
\end{equation}
\noindent
may be chosen as follows:
\begin{gather}
t, u, u_t, u_{tt},r=x^2+y^2, xu_x+yu_y, u_x^2+u_y^2, u_{xx}+u_{yy}, \label{DI-rotation}\\
u_x^2u_{xx}+2u_xu_yu_{xy}+u_y^2u_{yy}, xu_xu_{xx}+(xu_y+yu_x)u_{xy}+yu_yu_{yy}, \nonumber \\
u_{xt}^2+u_{yt}^2,  xu_{xt}+yu_{yt}, \nonumber \\
\frac{u_k}{x_k}, \frac{{u_k}_t}{x_k},\frac{u_{kl}}{x_k x_l} - \epsilon_{kl}\frac{u_k}{x_k^3} \label{CDI-rotation}
\end{gather}

We used notations $x_1=x, x_2=y$, $u_1=u_x, u_2=u_y$ etc.; $\epsilon_{kl}=1$ if $k=l$
or $\epsilon_{kl}=0$ if $k \ne l$.

Invariants (\ref{DI-rotation}) represent a functional basis of absolute differential
invariants for the operator (\ref{rotation})
(see e.g. \cite{FYeDifInvs}), and invariants (\ref{CDI-rotation}) are proper
conditional differential invariants under condition (\ref{cond-rotation}). It is easy to check
directly that they are really differential invariants under such condition.
The listed proper conditional differential invariants do not actually represent a functional
basis: e.g. from (\ref{cond-rotation}) $\frac{u_x}{x}=\frac{u_y}{y}$, but we adduced all such invariants
just to show their general structure.

The general form of the equation (\ref{eq1}) reducible with the ansatz
\begin{equation}
 u=\phi(t,r), \label{anz-r}
\end{equation}
\noindent
will be
\begin{equation}
 F(I_A,\frac{u_k}{x_k}, \frac{{u_k}_t}{x_k},\frac{u_{kl}}{x_k x_l} - \epsilon_{kl}\frac{u_k}{x_k^3})=0, \label{eq-r}
\end{equation}
\noindent
where $I_A$ is the functional basis of absolute differential invariants (\ref{DI-rotation}),
and the remaining variables are represented by proper conditional differential invariants.

It is easy to check that equation (\ref{eq-r}) can be reduced by means of the ansatz (\ref{anz-r}) to the form
\begin{equation}
 f(t,r,\phi,\phi_t,\phi_r,\phi_{tt},\phi_{tr},\phi_{rr})=0, \label{red-eq-r}
\end{equation}
\noindent 
and the class (\ref{red-eq-r}) may be studied to find equations having new symmetries.
Equations with hidden symmetries then will be described by conditional differential invariants under (\ref{cond-rotation})
and these new symmetries.

Reduction of equation (\ref{eq1}) by means of the new variable (\ref{rho})
is equivalent to its conditional invariance under the operators of Lorentz algebra
\begin{equation}
J_{01}=t\partial_x+x\partial_t, \qquad J_{02}=t\partial_y+y\partial_t,
\qquad J=x\partial_y-y\partial_x. \label{Lorentz}
\end{equation}

Conditional differential invariants with the conditions
\begin{equation}
tu_x+xu_t=0, \qquad tu_y+yu_t=0,
\qquad xu_y-yu_x=0. \label{cond-Lorentz}
\end{equation}
\noindent
may be chosen as follows:

\begin{gather}
u, x_\mu x_\mu, x_\mu u_\mu, u_\mu u_\mu, \Box u, u_\mu u_{\mu\nu}u_\nu, u_\mu u_{\mu\nu}u_{\nu\alpha}u_\alpha,   \label{DI-Lorentz}\\
u_{\mu\nu}u_{\nu\alpha}u_{\mu\alpha}, x_\mu u_{\mu\nu}u_\nu, x_\mu u_{\mu\nu}u_{\nu\alpha}u_\alpha, \nonumber \\
\frac{u_\mu}{x_\mu}, \frac{u_{\mu\nu}}{x_\mu x_\nu} - g_{\mu\nu}\frac{u_\mu}{x_\mu^3}.  \label{CDI-Lorentz}
\end{gather}

Here $\mu, \nu, \alpha$ take values from 0 to 2, and we used notations
$x_0=t, x_1=x, x_2=y$, $u_0=u_t,u_1=u_x, u_2=u_y$ etc.; $g_{\mu\nu}=(1,-1,-1)$.

Invariants (\ref{DI-Lorentz}) represent a functional basis of absolute differential
invariants for the operator (\ref{Lorentz}), and invariants (\ref{CDI-Lorentz}) are proper
conditional differential invariants under condition (\ref{cond-Lorentz}). It is easy to check
directly that they are really differential invariants under such condition.
The listed proper conditional differential invariants do not actually represent a functional
basis: e.g. from (\ref{cond-Lorentz}) $\frac{u_x}{x}=\frac{u_y}{y}$, but we adduced such invariants
just to show their general structure.

The general form of the equation (\ref{eq1}) reducible with the ansatz
\begin{equation}
 u=\phi(\rho), \label{anz-rho}
\end{equation}
\noindent
will be
\begin{equation}
 F(I_A,\frac{u_\mu}{x_\mu}, \frac{u_{\mu\nu}}{x_\mu x_\nu} - g_{\mu\nu}\frac{u_\mu}{x_\mu^3})=0, \label{eq-rho}
\end{equation}
\noindent
where $I_A$ is the functional basis of absolute differential invariants (\ref{DI-Lorentz}),
and the remaining variables are represented by proper conditional differential invariants.

It is easy to check that equation (\ref{eq-rho}) can be reduced by means of the ansatz (\ref{anz-rho}) to the form
\begin{equation}
 f(\rho,\phi,\phi',\phi'')=0, \label{red-eq-rho}
\end{equation}
\noindent
and the class (\ref{red-eq-rho}) may be studied to find equations having new symmetries.
Equations with hidden symmetries then will be described by conditional differential invariants under (\ref{cond-Lorentz})
and these new symmetries.

The presented results can be naturally extended to arbitrary number of space dimensions.

An example of a nonlinear wave equation having conditional symmetry with respect to the Lorentz group of the type (\ref{Lorentz}) with
$n$ space dimensions was given in \cite{FTs}:
\begin{equation}
\Box u = \frac{\lambda_0 u_0^2}{x_0^2}+\frac{\lambda_1 u_1^2}{x_1^2}+ \dots +\frac{\lambda_n u_n^2}{x_n^2}.  \nonumber
\end{equation}
\noindent
It is easy to see that this equation is actually constructed with first-order conditional
differential invariants of the type $\frac{u_\mu}{x_\mu}$.

\section{Conclusions}
We presented examples related to description of equations with
some specified conditional and hidden symmetry, as well as algorithms for group classification
of classes of equations with respect to conditional and hidden symmetry and possible reductions.

Further research in this direction shall involve classification with respect to Lie and conditional
symmetry of the remarkable classes of reduced equations.

\end{document}